# Study on the Data Processing of the IOT Sensor Network Based on Hadoop Cloud Platform and TWLGA Scheduling Algorithm

Guo-yu Li*, Kang Yang

**Abstract**
The Internet of Things (IOT) sensor network is an effective solution for monitoring environment condition. IOT sensor network generates massive data, and the abilities of massive data storage, processing and query become technical challenges. To solve the problem, a Hadoop cloud platform is proposed. With the help of time and workload genetic algorithm (TWLGA), the data processing platform provides the work of one node to share with others, which not only raises efficiency of one single node, but also provides the compatibility support to reduce the possible risk of software and hardware. In the experiment, a Hadoop cluster platform with TWLGA scheduling algorithm is built, and the performance of the platform is tested. The results show that the Hadoop cloud platform is suitable for big data processing of the IOT sensor network.

## 1. Introduction

The IOT allows sensor network to constantly sense its environmental condition for relevant movements and initiate it based on preprogrammed rules [1-3]. Hence, the sensor network is the basis of the IOT. With the expansion of the IOT sensor network, various intelligent sensors have been used in many applications, which makes the ability of the high-speed data acquisition and processing to be the key issue for IOT sensor network. Since single node contains plenty sensors working at the same time, real-timely processing of the mass data is exceedingly required. Besides, the managed node removes redundant data based on judgment. If all the work is done by one single node, that node will be abnormally busy and fail to respond in time, which will lead the paralysis of whole network. In order to process massive data real-timely, the cloud computing for IOT network arises at the time [4,5]. The cloud computing provides a usable way to process the resource: the cloud computing divides the whole computing processes across a computer network into many subroutines and distributes these subroutines to many spare servers across the network. After calculation and analysis, the results are given back to the user. The cloud computing is characterized by the mass data storage, the high-speed data analysis and the real-time processing [6,7]. Hadoop is a distributed computing framework, which can run applications on a large number of low-cost hardware devices in a cluster. The Hadoop-based framework is used for big data collection, processing and storage in air pollution monitoring, fault detection and disaster management [8,9]. In this paper, the fiber Bragg grating sensors and traditional sensors form a sensor network to collect the temperature data continuously, the temperature data is continuously collected, stored and accumulated by the computers, and finally forms different sizes of packets. To enhance the ability of real-time data processing of the sensor network, the cloud-computing-based data processing platform for sensor network is investigated. Furthermore, the performance of the Hadoop cluster platform with TWLGA is studied.

## 2. IOT Sensor Network Based on Cloud Computing

For IOT sensor network, when a node of IOT sensor network collects and processes massive data in real time, the node will become busy and fail to respond in time. To avoid paralysis, the data is distributed from one busy node to other idle nodes by the cloud computing technology. These nodes share resources with one managed node. In this way, data processing is not concentrated in busy nodes, but is distributed over many idle nodes. All the calculated results are given back to the managed node. Therefore, cloud computing technology not only enhance the real-time processing ability of the nodes, but also ensures the high-speed acquisition and analysis of massive data. As the resources are calculated by a series of idle nodes instead of the busy nodes, cloud computing technology reduces the risk of computing workload. Therefore, the cloud-computing-based IOT sensor network is more reliable, manageable and flexible.

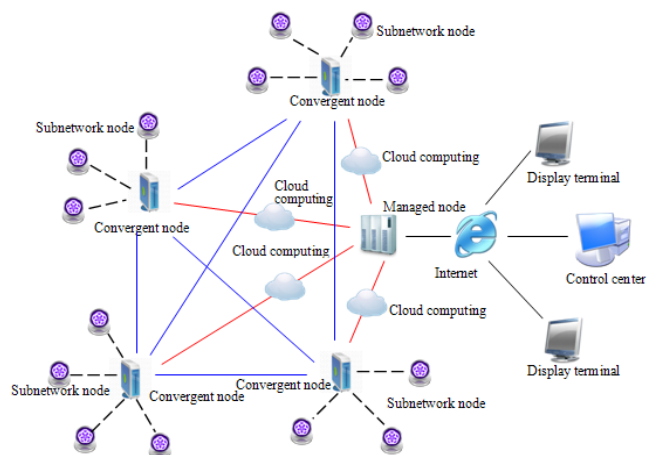

**Fig. 1.** The IOT sensor network based on cloud computing.

The data processing of the IOT sensor network is constantly utilized by using the theoretical model of cloud computing. The overall technical schematic diagram of IOT sensor network based on cloud computing is shown as Fig. 1. In the Fig.1, many subnetworks consist the IOT sensor network, and each subnetwork contains a series of sensors. The sensing elements of the subnetworks monitor the variation of parameters. The convergent nodes are connected each other, and each convergent node connects to many subnetworks. These nodes are all connected to the managed nodes by the cloud computing technology. Based on information of the managed node, the data resources are calculated by a series of idle nodes instead of the busy nodes. The managed nodes are controlled by control center through commands over the internet. The display terminal is used to display the analysis results of IOT sensor network.

## 3. Time and Workload Genetic Algorithm

Generally, a Hadoop cluster is composed of hardware computers and software, and the scheduler completes the specific assignment of tasks. Hadoop uses the FIFO (first in, first out) scheduler by default. FIFO scheduling algorithm is a queue form algorithm, it is relatively simple, and can not meet the actual complex requirements. In the experiment, the Hadoop cluster is built by the computers with different configurations, computing power and workload, so some tasks will be improperly assigned to nodes with heavy workload. In order to avoid this problem, TWLGA considering the dual constraints of time and workload is proposed, which enables clusters to meet the requirements of short task running time and more reasonable assignment [10,11].

The TWLGA task scheduling includes chromosome coding [12], node workload [13], fitness function [14], crossing and variation [15].

## 3.1 Chromosome Coding

Data processing adopts the indirect encoding of resource tasks. Firstly, the number of task slices and the resource number of corresponding task slices are determined, and then the chromosome coding is completed. Secondly, the chromosomes are decoded, and the task-resource tables are attained. Thirdly, the expect time to complete (ETC) matrix can be used to calculate the time of each resource which completes all the tasks assigned to it. Assuming that the subtasks assigned to resource $i$ are $M_i$, then the time required for resource i to complete all subtasks is

$$EachResourceTime = \sum_{j=1}^{n} Time(i,j), i \in [1, M] \tag{1}$$

Where, $n$ represents the number of subtasks of each resource, and $Time(i,j)$ represents the time to execute the jth task of the resource i. In cloud computing, the tasks are all computed in parallel, so the one with longest running time of the all resources can be thought as the final completion time of the task.

$$JobFinalTime(J) = max(EachResouceTime(i)) \tag{2}$$

## 3.2 Node Workload

In Hadoop cluster, the three major categories of computer roles are client computer, master node and slave nodes. The role of client computer is to load data to the cluster and submit it to the MapReduce. The master node has the two key functional pieces: storing lots of data and running parallel computations on all that data. The slave nodes make up the vast majority of computers and do all the dirty work of storing the data and running the computations. Each slave runs both a data node and task tracker daemon that communicate with and receive instructions from their master node. The workload of each slave node is affected by a number of factors, including CPU usage, memory usage and network resources. In Hadoop cluster, only four factors are usually considered. Define the workload of the nodes as following formula:

$$WorkLoad(i) = W_{cpu} \times \mu_{cpu} + W_{me} \times \mu_{me} + W_{disk} \times \mu_{disk} + W_{nr} \times \mu_{nr} \tag{3}$$

Where, $W_{cpu}$, $W_{me}$, $W_{disk}$ and $W_{nr}$ represent the proportion of CPU, memory, disk and network resources of the nodes in the overall performance. $\mu_{cpu}$, $\mu_{me}$, $\mu_{disk}$ and $\mu_{nr}$ represent the usage of the CPU, memory, disk and network resources respectively.

In the experiment, the master computer and slave computers collect and store data from the sensor network, but some slave nodes have a heavy workload and other slave nodes are relatively idle, so the master computer provides the work of one node to share with others, which not only raises efficiency of one single node, but also provides the compatibility support to reduce the possible risk of software and hardware.

## 3.3 Fitness Function

Assuming that the initial population is $P$, the number of computational resources is $R$, and the number of sub-tasks is $N$, the random coding of the chromosomes is generated by random initialization, that is to say, a total of $P$ chromosomes, whose length is R, and the number gene is randomly selected in $[1, R]$.

The selection of fitness function directly not only determines the quality of the algorithm results, but also directly affects the speed of the task running time and the appropriateness of node allocation. From the above analysis, it is well known that the fitness based on GA considering the running time is $JobFinalTime(J)$, so the fitness function can be established as

$$P_{time} = \frac{1}{JobFinalTime(J)} \qquad (4)$$

However, in the fitness function, only the task running time is considered, the workload of the assigned node itself does not consider, so the workload factor is taken into account and the fitness function is improved

$$Optimum(J) = \frac{1}{JobFinalTime(J) \times (1 + WorkLoad(i))} \qquad (5)$$

In this way, the workload and the task scheduling are all considered, so the problem that the tasks are assigned to nodes with heavy workload is avoided.

## 3.4 Crossing and Variation

The function of crossover is to generate different individuals through the crossover transformation of the genes, which is the important basis of the whole algorithm and determines the validity of the following operations. The mutation operation can maintain and improve the diversity of species to improve the local search ability. The TWLGA improves the probability formula of the cross variation, so that they can achieve adaptive adjustment. The improved probability formula is as follows

$$P_c = \frac{P_{c1} - P_{c2}}{P_{c1}} \times \frac{f_0 - f'}{f_0 - \bar{f}} \qquad (6)$$

$$P_m = \frac{P_{m1} - P_{m2}}{P_{m1}} \times (f_0 - f') / (f_0 - \bar{f}) \qquad (7)$$

Where, $f$ is the fitness of the mutant individual, $f_0$ is the maximum fitness of the population, $f'$ is the maximum fitness of the crossover individual, and $\bar{f}$ is the average fitness of the population. After the improvement of the sum calculation formula $P_c$ and $P_m$, it can complete the task scheduling well, reduce the total time of the task execution, and take the workload of the nodes into account, which makes the calculation efficiency greatly improved.

# 4. Experimental Results and Test

The Hadoop cluster platform is built to test the performance of data processing system based on cloud computing with TWLGA. The experimental setup is shown in Fig. 2, the Hadoop cluster is composed of five computers, one is acted as the master computer, and the others are acted as the slave computers. The name of five computers is changed before Hadoop software installation, so the name of five computers is changed to master, slave1, slave2, slave3 and slave4 separately in the catalog /etc/hostname of each computer. Then the hostname and the IP address are added to the configuration file in the catalog /etc/hosts, so each node computer can be recognized and accessed each other.

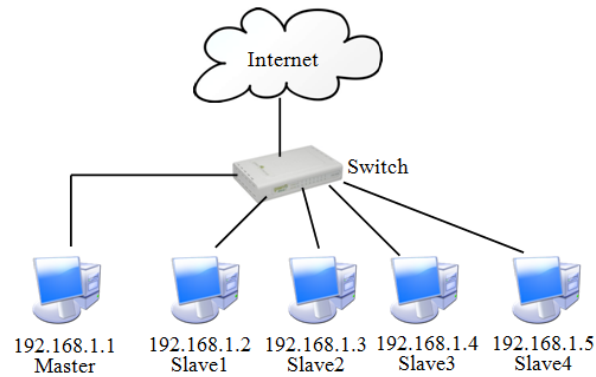

**Fig. 2.** The network topology of Hadoop cluster.

The operating system of each node computer is ubuntu 12.04.3, and the version of Hadoop software is 1.0.2. Hbase is acted as the database of the Hadoop cluster, whose worksheets are used to store the sensor data. The version of Hbase is 0.94.10 and the Hbase is deployed to the cloud computing center. The tomcat used in the cluster is a web server and it is installed in the master computer, which is deployed by the myeclipse software.

Table. 1 shows the source data format collected from a fiber sensor, the data respectively represents year, month, day, observation time (hour and minute), wavelength and so on. The temperature is calculated from wavelength data. There is a lot of data and the format is complex. According to the demand, users only care about the year, month, time and corresponding temperature, so they only need to take out the part of the data. Since there are thousands of data files, and the advantage of Hadoop is processing large files. The processing efficiency of small files is very low, so it is not advisable to process these small files directly. In the experiment, the file merging function of Hadoop is used to merge all files of the same year. In this way, a single large file is created for each year, and the useful data is extracted by MapReduce, which takes full advantage of Hadoop's ability to handle large files.

**Table 1.** The Source data format

| Year | Month | Day | Hour | Minute | Wavelength |
|------|-------|-----|------|--------|------------|
| 2021 | 03 | 01 | 08 | 30 | 154574 |
| 2021 | 03 | 01 | 08 | 30 | 154577 |
| 2021 | 03 | 01 | 08 | 31 | 154575 |
| 2021 | 03 | 01 | 08 | 31 | 154579 |
| --- | --- | --- | --- | --- | --- |
| 2021 | 03 | 01 | 11 | 20 | 154593 |
| 2021 | 03 | 01 | 11 | 20 | 154593 |

## 4.1 Hadoop I/O Performance Test

As enhancing the response speed and processing power is the main challenges for the data processing system, the performance test of the Hadoop I/O is necessary. The reading and writing speed test of the Hadoop I/O are carried out by the hadoop-1.0.2.jar. The reading and writing speed of the Hadoop Distributed File System (HDFS) are tested through the MapReduce task, so the 10 512M files are read and written to test the HDFS I/O performance, the experimental test results are shown in Table. 2.

**Table 2.** The HDFS I/O performance test

| Symbol | Read | Write |
|---|---|---|
| Number of files | 10 | 10 |
| Total Mbytes processed | 5120 | 5120 |
| Throughput(Mb/sec) | 17.72 | 3.49 |
| Average I/O rate (Mb/sec) | 114.20 | 4.14 |
| I/O rate std deviation | 120.74 | 1.92 |
| Test exec time(sec) | 147.08 | 313.39 |

From the Table. 2, the writing speed of the Hadoop cluster is 4.14Mb/sec, but the reading speed is 114.20Mb/sec, the reading speed is about 30 times faster than the writing speed, so the Hadoop cluster is mainly used for reading operation, especially, it is suitable for one time writing and reading multiple times.

## 4.2 MapReduce Performance Test

MapReduce is an important programming model for large-scale data parallel and distributed application. And Hadoop is an associated implementation of MapReduce with open-source. The MapReduce performance test files are 5 txt files, the sizes are 160Mb, 320Mb, 640Mb, 1.3Gb and 2.6Gb separately. The five files are placed and counted the one node, two nodes and three nodes, the experimental test results are shown in the Table. 3.

**Table 3.** The MapReduce performance test

| Symbol | 160Mb | 320Mb | 640Mb | 1.3Gb | 2.6Gb |
|---|---|---|---|---|---|
| 1node(sec) | 29 | 64 | 408 | 548 | 1213 |
| 2nodes(sec) | 38 | 90 | 359 | 487 | 1054 |
| 3nodes(sec) | 55 | 116 | 352 | 453 | 901 |

From the Table.3, the results show that, in the case of small amount of data, the more nodes it has, the slower the calculation speed is. As the amount of data increases, the multi-nodes system embodies the superiority. When dealing with small amount of data, the MapReduce must read the data from all of the nodes, so the time spent on network transmission is also essential. Once large-scale data need to be processed, the role of MapReduce is crucial, which fully demonstrates that MapReduce is suitable for processing the big data.

## 4.3 HBase Performance Test

In order to test the HBase performance, the general performance testing tool from Yahoo is used. The writing time, data throughput and the reading time test are carried out separately when the threads are 1, 50, 100, 1000 and 5000. The test results are shown in the Fig.3, Fig.4 and Fig.5.

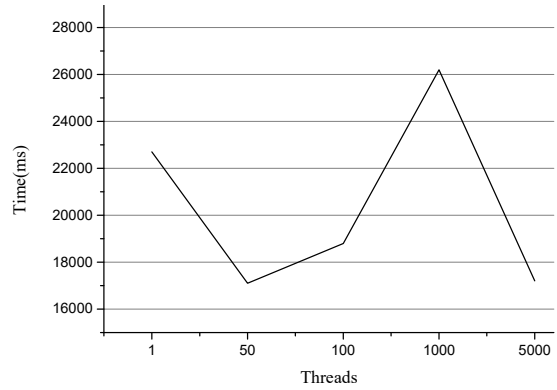

**Fig. 3.** The writing time change with different threads.

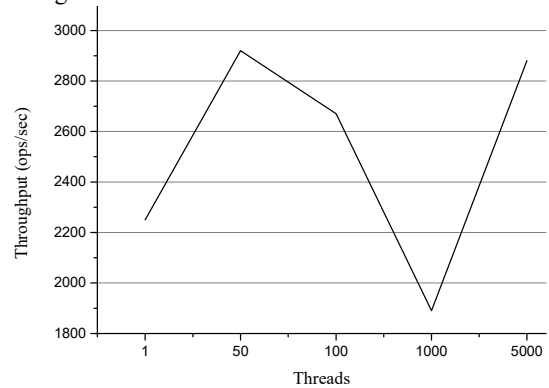

**Fig. 4.** The throughput change with different threads.

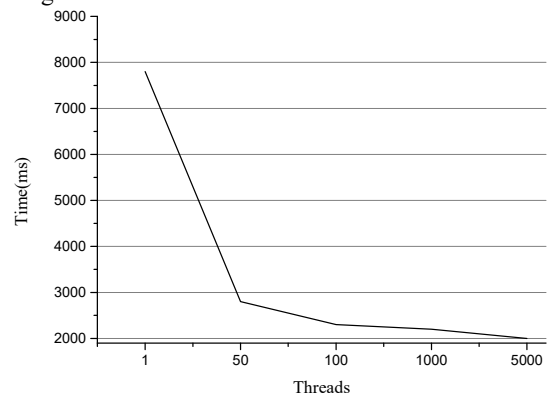

**Fig. 5.** The reading time change with different threads.

Fig.3 and Fig.4 show the HBase performance test when writing data, the writing time and data throughput are changing inversely as the number of threads increases. Fig.5 shows the HBase performance test when reading data, the reading time decreases exponentially and then flattens out with the increase of threads.

From the figures, whether writing or reading data, the time of processing data does not increase in the case of the increasing number of threads, the throughput is within the range of 2300 to 2900, and there is no significant change in overall. So the HBase still has fast processing speed under the condition of the high concurrency.

# 5. Conclusions

In this work, the data processing of IOT sensor network based on Hadoop cloud platform and TWLGA scheduling algorithm is proposed. For the sake of improving the platform performance, the cloud computing technology with TWLGA is adopted to process massive data and avoid paralysis of IOT sensor network. The workload of single node is shared to others. In this way, the efficiency of a single node is enhanced and the possible risk of the network is also reduced. Finally, a Hadoop cluster platform is built, and the performance of the platform is tested. The results show that the Hadoop cluster platform is suitable for big data processing of the IOT sensor network.

# Acknowledgement

This paper is supported by the project of National Natural Science Foundation of China (No.62175055) and the Research Fund of Handan University (No. 16215).

# References

[1] A. Sarma, J. Girao, "Identities in the Future Internet of Things," *Wireless Pers Commun.,* vol.49, no.3, pp.353-363, Jan.2009.

[2] Debasis Bandyopadhyay Jaydip Sen, "Internet of Things:Applications and Challenges in Technology and Standardization," *Wireless Pers Commun.,* vol.58, no.1, pp.49–69, Apr. 2011.

[3] R. Roman, "Key Management Systems for Sensor Networks in the Context of the Internet of Things," *Computers & Electrical Eng.,* vol.37, no.2, pp.147-159, Mar. 2011.

[4] Lizhe Wang, Gregor von Laszewski, Andrew Younge, "Cloud Computing: a Perspective Study," *New Generation Computing,* vol.28, no.2, pp.137-146, Apr. 2010.

[5] Yan, J., Wang, X., Gan, Q., et.al, "Secure and efficient big data deduplication in fog computing," *Soft Computing,* vol.24, no.8, pp.5671-5682, Jul. 2020.

[6] P. Liu, *Cloud Computing,* Electronic Industry Press Beijing, 2010.

[7] Yedidsion, H., Ashur, S., Banik, A., et al., "Sensor network topology design and analysis for efficient data gathering by a mobile mule," *Algorithmica,* vol.82, no.10, pp.2784-2808, Oct. 2020.

[8] Suganya E., Rajan C., "An adaboost-modified classifier using particle swarm optimization and stochastic diffusion search in wireless IoT networks," *Wireless Networks*, Vol.4, pp.1-13, Nov. 2020.

[9] Preethi, K., and R. Tamilarasan, "Monitoring of air pollution to establish optimal less polluted path by utilizing wireless sensor network." *Journal of Ambient Intelligence and Humanized Computing,* Vol.12, pp.6375-6386, Jun. 2020.

[10] Yadav P K , "Workload Analysis in a Grid Computing Environment: A Genetic Approach,", *International Journal of Computer Applications,* vol.93, no.16, pp.26-29, May. 2014.

[11] Aziza, H., & Krichen, S., "A hybrid genetic algorithm for scientific workflow scheduling in cloud environment," *Neural Computing and Applications*, vol.32, no.12, pp.1-16, Sep. 2020.

[12] Tuncer A , Yildirim M., "Chromosome Coding Methods in Genetic Algorithm for Path Planning of Mobile Robots," *Computer and information Sciences Ⅱ*, pp.377-383, Sep. 2011.

[13] Li, L., Guo M., Ma L., et al., "Online Workload Allocation via Fog-Fog-Cloud Cooperation to Reduce IoT Task Service Delay," *Sensors*, vol.19, no.18, pp.3830, Sep. 2019.

[14] Shah, S., Rashid, M., Arif, M., "Estimating WCET using prediction models to compute fitness function of a genetic algorithm," *Real-Time Syst.* vol.56, pp.28–63, Feb. 2020.

[15] Zhou, Z., Li, F., Zhu, H. et al., "An improved genetic algorithm using greedy strategy toward task scheduling optimization in cloud environments," *Neural Comput & Applic.,* vol.32, pp.1531–1541, Nov.2020.

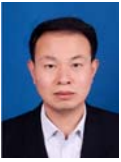

**Guoyu Li**   https://orcid.org/0000-0003-1735-0536  (ORCID ID)

He received B.S. degree in physics from Hebei Normal University in 2001, M.S. degrees in microelectronics and solid-state electronics from Hebei University of Technology in 2004, and Ph.D. degree in optics from Nankai University in 2007. He is currently a professor in the School of Information Engineering, Handan University, Handan, China. His current research interests include sensor network and IoT.

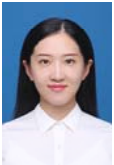

**Kang Yang**   https://orcid.org/0000-0002-0452-7868  (ORCID ID)

She received B.S. degree in electronics science and technology from Hebei University of Technology in 2007, M.S. degree in communication and information systems from Hebei University of Technology in 2010, and Ph.D. degree in optics from the Nankai University in 2019. She is currently an associate professor in the School of Information Engineering, Handan University, Handan, China. Her research interests include communication and communication network.